\documentclass{article}

\PassOptionsToPackage{square,sort&compress}{natbib}

 \usepackage[dblblindworkshop,final]{neurips_2025}
\workshoptitle{Machine Learning and the Physical Sciences}

\usepackage[utf8]{inputenc} 
\usepackage[T1]{fontenc}    
\usepackage{hyperref}       
\usepackage{url}            
\usepackage{booktabs}       
\usepackage{amsfonts}       
\usepackage{nicefrac}       
\usepackage{microtype}      
\usepackage[dvipsnames]{xcolor}         
\usepackage{siunitx}
\usepackage{CJKutf8}
\usepackage{tikz}
\usepackage{caption}
\usepackage{subcaption}
\usepackage{wrapfig}
\usepackage{titlesec}
\usepackage{amsmath,amssymb}
\usepackage[skip=2pt]{parskip}
\usetikzlibrary{positioning, arrows.meta, shapes.multipart}

\title{Universal Spectral Tokenization via Self-Supervised Panchromatic Representation Learning}

\author{
Jeff Shen\thanks{shenjeff@princeton.edu}\textsuperscript{\,\,\,\,,1},
Francois Lanusse\textsuperscript{2,3},
Liam Parker\textsuperscript{6,3,4,5},
Ollie Liu\textsuperscript{7},
Tom Hehir\textsuperscript{8}
\And
Leopoldo Sarra\textsuperscript{3},
Lucas Meyer\textsuperscript{3},
Micah Bowles\textsuperscript{9},
Sebastian Wagner-Carena\textsuperscript{3,5}
\And
Helen Qu\textsuperscript{3},
Siavash Golkar\textsuperscript{3,5},
Alberto Bietti\textsuperscript{3},
Hatim Bourfoune\textsuperscript{10},
Nathan Cassereau\textsuperscript{10}
\And
Pierre Cornette\textsuperscript{10},
Keiya Hirashima\textsuperscript{3,11},
Geraud Krawezik\textsuperscript{3},
Ruben Ohana\textsuperscript{3},
Nicholas Lourie\textsuperscript{5}
\And
Michael McCabe\textsuperscript{3,5},
Rudy Morel\textsuperscript{3},
Payel Mukhopadhyay\textsuperscript{1,8},
Mariel Pettee\textsuperscript{12}
\And
Bruno Regaldo-Saint Blancard\textsuperscript{3},
KyungHyun Cho\textsuperscript{5},
Miles Cranmer\textsuperscript{8},
Shirley Ho\textsuperscript{1,3,5}
}
\date{}

\usepackage{hyperref}
\hypersetup{
    colorlinks = true,
    linkbordercolor = white,
    citecolor = MidnightBlue,
    linkcolor = MidnightBlue,
}
\begin{document}
\begin{CJK*}{UTF8}{gbsn}

\maketitle
\vspace{-0.9cm}

\centerline{The Polymathic AI Collaboration}
\vspace{0.3cm}

\centerline{\begin{minipage}{0.8\textwidth}
\textsuperscript{1}Princeton University,
\textsuperscript{2}Université Paris-Saclay, Université Paris Cité, CEA, CNRS, AIM,
\textsuperscript{3}Flatiron Institute,
\textsuperscript{4}Lawrence Berkeley National Laboratory,
\textsuperscript{5}New York University,
\textsuperscript{6}University of California, Berkeley,
\textsuperscript{7}University of Southern California,
\textsuperscript{8}University of Cambridge,
\textsuperscript{9}University of Oxford,
\textsuperscript{10}IDRIS, CNRS,
\textsuperscript{11}RIKEN Center for iTHEMS,
\textsuperscript{12}University of Wisconsin–Madison
\end{minipage}}
\vspace{0.2cm}

\begin{abstract}
Sequential scientific data span many resolutions and domains, and unifying them into a common representation is a key step toward developing foundation models for the sciences.
Astronomical spectra exemplify this challenge: massive surveys have collected millions of spectra across a wide range of wavelengths and resolutions, yet analyses remain fragmented across spectral domains (e.g., optical vs. infrared) and object types (e.g., stars vs. galaxies), limiting the ability to pool information across datasets.
We present a deep learning model that jointly learns from heterogeneous spectra in a self-supervised manner. 
Our universal spectral tokenizer processes spectra from a variety of object types and resolutions directly on their native wavelength grids, producing intrinsically aligned, homogeneous, and physically meaningful representations that can be efficiently adapted to achieve competitive performance across a range of downstream tasks.
For the first time, we demonstrate that a single model can unify spectral data across resolutions and domains, suggesting that our model can serve as a powerful building block for foundation models in astronomy---and potentially extend to other scientific domains with heterogeneous sequential data, such as climate and healthcare.
\end{abstract}

\section{Introduction}

Spectra encode fundamental astrophysical information, from stellar chemical abundances to galaxy dynamics and the state of the intergalactic medium.
Large-scale surveys such as SDSS, DESI, GALAH, and APOGEE \citep{sdss_dr17, desi_dr1, galah_dr3} have collected millions of spectra across a wide range of wavelengths ($R\sim 2{,}000$–$28{,}000$) and science targets.
However, analysis pipelines remain fragmented: each survey typically requires bespoke preprocessing and task-specific machine learning (ML) models.
This siloed approach limits flexibility, prevents the reuse of knowledge across datasets, and makes it impossible to combine heterogeneous information into a shared representation.
Moreover, training a new model for every instrument, object class, or task is inefficient, hinders generalization, and stands in contrast to the foundation model paradigm that is beginning to take hold in astronomy \citep[e.g.,][]{Rizhko2024, smith2024astropt, parker2024astroclip}.
Beyond astronomy, this challenge reflects a broader problem: how to learn universal representations from irregular, multiresolution sequential data---a setting that also arises in time series across physics, climate, and healthcare.

We propose a universal tokenizer for spectra that directly ingests native wavelength grids without resampling, enabling seamless integration across surveys.
Because the architecture is resolution- and domain-agnostic, the same design naturally extends beyond spectra; for example, wavelength embeddings can be swapped for time-based embeddings to tokenize irregular, multiresolution time series.
This positions our approach as a flexible building block for scientific foundation models, in astronomy and more broadly.

\paragraph{Related work}
Physically motivated spectral models fit forward models to spectra, offering interpretability but suffering from model misspecification and poor scalability. 
On the data-driven side, supervised approaches achieve high accuracy where labels exist, but are survey-specific and limited by label availability.
\cite{koblischke2024spectrafm} show, using a similar encoder to the one we propose, that a foundation model approach (pretraining, then fine-tuning) can reduce the limitations of label availability on a stellar parameter regression problem using APOGEE spectra.
Self-supervised spectral models \citep[e.g.,][]{Portillo2020, Melchior2022} have been explored in single domains on fixed (latent) grids, which introduce interpolation artifacts and scale poorly across wide wavelength ranges.
Furthermore, they typically require a good estimate of the redshift before processing.\footnote{This can be problematic as obtaining a good estimate of the redshift requires a good spectral model, thus leading to a chicken-and-egg problem.}
In contrast, our approach requires nothing but the spectrum.
Contrastive methods \citep{buck2024deep, zhao2025specclip} learn spectral representations but still rely on survey-specific encoders/decoders. 
In contrast, our approach scales to arbitrary datasets with a single encoder and decoder.

\paragraph{Contributions}
To our knowledge, no prior work has demonstrated a single model that can pretrain jointly across heterogeneous spectroscopic surveys (optical and IR; low- and high-resolution; stars, galaxies, and quasars) on their native wavelength grids without homogenization, while also providing an architecture that naturally extends to other forms of irregular, multiresolution sequential data such as time series.
Our contributions are:
\textbf{(1)} a novel architecture that directly ingests sequential data on arbitrary combinations of wavelength/time grids, 
\textbf{(2)} a panchromatic, multi-resolution, self-supervised pretraining strategy applied to SDSS, DESI, GALAH, and APOGEE, yielding cross-domain, homogeneous, and rich representations under a single model, and
\textbf{(3)} demonstration that lightweight adaptation of these embeddings achieves competitive performance on downstream tasks such as object classification and stellar parameter regression relative to task-specific baselines.

\section{A Universal Spectrum Tokenizer}

Our model is based on a Vision Transformer (ViT) architecture \citep{Dosovitskiy2021}, adapted for one-dimensional spectral data. 
In particular, our innovative encoder creates homogeneous, wavelength-aware embeddings from heterogeneous input spectra, allowing these embeddings to be used for a variety of downstream tasks. 
Furthermore, the architecture allows the model to be easily extended to pretraining with a masked autoencoding or contrastive objective. 
A diagram of the overall architecture and pretraining process is shown in Figure \ref{fig:model_architecture}.

\newcommand\rotnode[5]{%
  \node [#1, opacity=1.0] (#3) {#4};
  \node [#2, rotate around={#4:(#3.center)}] at (#3) (#3_box) {#5};
}
    
\begin{figure}[t]
\centering
\begin{tikzpicture}[
  layer/.style={rectangle, draw=black, fill=#1!20, minimum width=2.1cm, minimum height=0.6cm, align=center, font=\small},
  arrow/.style={->, thick},
  every node/.style={font=\small}
]

    \rotnode{}{layer=gray}{patch}{270}{Normalize\\Patch\\Project};
    \rotnode{right=0.5cm of patch}{layer=Periwinkle}{encoder}{270}{Encoder};
    \rotnode{right=0.8cm of encoder}{layer=Cerulean}{embeddings}{270}{Homogeneous\\embeddings};
    \rotnode{right=0.8cm of embeddings}{layer=Periwinkle}{decoder}{270}{Decoder};
    \rotnode{right=0.5cm of decoder}{layer=gray}{unpatch}{270}{Unnormalize\\Unpatch\\Deproject};

    \node[inner sep=0pt, left=1cm of patch] (spec_in) {\includegraphics[width=.2\textwidth]{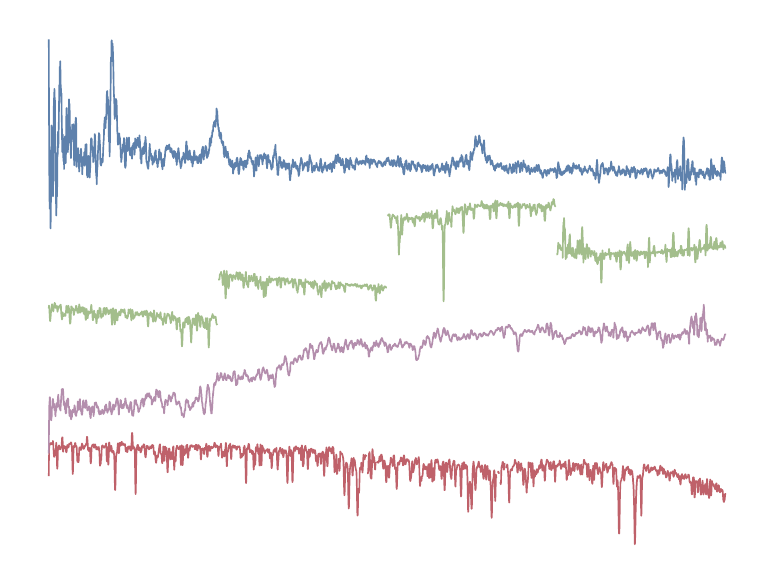}};

    \node[inner sep=0pt, right=1cm of unpatch] (spec_out) {\includegraphics[width=.2\textwidth]{figures/spectra_examples.pdf}};

    \node[layer=YellowGreen, above=1.5cm of embeddings, xshift=3.5cm] (classification) {Classification/regression};
    \node[layer=YellowGreen, above=1.5cm of embeddings, xshift=-3.3cm] (search) {Similarity search};
    \node[layer=YellowGreen, above=1.5cm of embeddings] (foundation) {Foundation model\\training};

    \node[below=0.3cm of spec_in] (wavelength) {Wavelength};

    \draw[arrow] (spec_in) -- (patch_box);
    \draw[arrow] (patch_box) -- (encoder_box);
    \draw[arrow] (encoder_box) -- (embeddings_box);
    \draw[arrow] (embeddings_box) -- (decoder_box);
    \draw[arrow] (decoder_box) -- (unpatch_box);
    \draw[arrow] (unpatch_box) -- (spec_out);

    \coordinate (emb_anchor) at ([yshift=0.18cm]embeddings_box.west);

    \draw[arrow] (embeddings_box.west) -| (emb_anchor) -| (classification.south);
    \draw[arrow] (embeddings_box.west) -| (emb_anchor) -| (search.south);
    \draw[arrow] (embeddings_box.west) -| (emb_anchor) -| (foundation.south);
    
    \draw[arrow] (wavelength) -| (patch_box.east);
    \draw[arrow] (wavelength) -| (decoder_box.east);

\end{tikzpicture}
\caption{Overview of the universal spectrum tokenizer. The encoder creates homogeneous, wavelength-aware embeddings from heterogeneous input spectra; these can be used for a variety of downstream tasks. The decoder reconstructs the input spectrum from the learned representations.}
\label{fig:model_architecture}
\vspace{-0.5cm}
\end{figure}

\paragraph{Input processing and encoding}


Each spectrum provides flux values with measurement errors, per-pixel wavelengths, and a mask for problematic pixels. 
We normalize spectra by dividing out the median flux, ensuring the model focuses on relative variations. 
The normalized flux and error values are then segmented into fixed-size patches, projected into a high-dimensional embedding space.
We also construct a patch-level mask that excludes patches dominated by bad or padded pixels (e.g., due to detector failures): if at least half of the pixels within a patch are bad, we mark the patch as bad.

To handle non-uniform wavelength grids, we add continuous per-pixel sinusoidal embeddings \citep{Rozanski2023}, 
avoiding the interpolation errors and inefficiencies of fixed latent grids \citep[e.g.,][]{Melchior2022}. 
For wavelength $\lambda$ and frequency scale $\omega_k$, the embedding is
\begin{align}
\text{PE}(\lambda)_k =
\begin{cases}
\sin(\omega_k \lambda), & k \ \text{even}, \\
\cos(\omega_k \lambda), & k \ \text{odd},
\end{cases}
\quad 
\omega_k \in \Big[ \tfrac{2\pi}{\text{max\_period}}, \tfrac{2\pi}{\text{min\_period}} \Big],
\end{align}
with $\{\omega_k\}$ log-spaced between $\tfrac{2\pi}{\text{max\_period}}$ and $\tfrac{2\pi}{\text{min\_period}}$. 
Wavelength embeddings are patched and added to the flux patches, imbuing the tokens with wavelength information.
These representations are processed by a series of transformer blocks \citep{Vaswani2017}, which consist of alternating multi-head self-attention and feed-forward layers.
Crucially, the attention mechanism allows the model to learn long-range dependencies between different parts of the spectrum, such as correlated line features that span large wavelength ranges.
This is particularly important for spectral data, where features can span large wavelength ranges and may not be immediately adjacent in the input sequence.
Bad patches are ignored in the attention computation according to the patch mask.
The output is, for each spectrum, a sequence of tokens with a sequence dimension (with length equal to the original sequence length divided by the patch size) and a small channel dimension, representing wavelength-aware spectral features; these can be used for various downstream tasks such as classification, regression, or reconstruction.

\paragraph{Decoding and reconstruction}
For self-supervised pretraining, we use a loss-aware reconstruction objective, which encourages the model to reconstruct the original input spectrum from the learned representations.
The decoder mirrors the encoder; the decoding process proceeds by first requesting an output wavelength grid, and adding its sinusoidal embedding to the encoder tokens to provide the model with information about where we want to reconstruct the signal.
The decoder then processes the tokens to produce a sequence of outputs, which are unpatched to produce a per-pixel reconstruction of the input spectrum.
The reconstruction loss is a Gaussian likelihood 
between the reconstructed spectrum and the original input spectrum 
and is only computed for valid pixels:
$
    \mathcal{L} = \frac{1}{N} \sum_{i=1}^{N} m_i \frac{(y_i - \hat{y}_i)^2}{\sigma_i^2},
$
where $N$ is the number of valid pixels, and $m_i, y_i, \hat{y}_i$, and $\sigma_i$ are the validity mask, original flux, reconstructed flux, and measurement error at pixel $i$.

\paragraph{Dataset}
We train our model using spectroscopic data from four major surveys: SDSS DR17, GALAH DR3, DESI DR1, and APOGEE; the data are accessed through the Multimodal Universe \citep{multimodal_universe} project.
We provide details about the datasets in Table \ref{tab:datasets};
we emphasize that our model is trained across all of these datasets without homogenization, efficiently leveraging the native wavelength grids and resolutions of each survey.
Doing the same with a fixed grid model such as the one in \cite{Melchior2022} would require a grid size of 300K pixels---an impossibility.

\begin{table}[t]
\centering
\footnotesize
\caption{Summary of spectroscopic datasets used for training.}
\label{tab:datasets}
\begin{tabular}{llll}
\toprule
\textbf{Dataset} & \textbf{Wavelength Range} & \textbf{Resolution} & \textbf{Objects} \\
\midrule
  SDSS DR17 \citep{sdss_dr17}        & 3600–10400 Å             & $R\sim2000$         & galaxies, quasars, stars \\ 
  DESI DR1 \citep{desi_dr1}        & 3600–9800 Å              & $R\sim5000$         & galaxies, quasars, stars \\
  GALAH DR3 \cite{galah_dr3}       & 4700-7900  Å             & $R\sim28,000$       & stars \\ 
  APOGEE \citep{sdss_dr17}          & 1.51–1.7 \si{\micro\meter} & $R\sim22,500$       & stars \\ 
\bottomrule
\end{tabular}
\vspace{-0.2cm}
\end{table}

\paragraph{Training and implementation details}
We use 6 encoder and 6 decoder transformer blocks, an embedding dimension of 512, 8 attention heads, and a patch size of 32 pixels.
We use a batch size of 64, a constant learning rate of 1e-4, and the AdamW optimizer \citep{adamw} with $\beta_1=0.9$, $\beta_2=0.999$, and weight decay of 0.01.
We train for 600k steps on 4 NVIDIA A100-SXM4-40GB GPUs, which takes 48 hours.

\section{Results}

This section shows that our model adapts to diverse tasks with minimal training, capturing information across spectral domains and object types. 
Our aim here is not to achieve state-of-the-art performance, but to demonstrate broad applicability. 
We leave multi-modal and cross-survey tasks to future foundation models built on top of our universal tokenizer, while emphasizing that, unlike task-specific pipelines, our model is reusable across many settings.

\paragraph{Spectrum reconstruction}
\begin{figure}[t]
\centering
\begin{subfigure}{0.48\textwidth}
    \includegraphics[width=\textwidth]{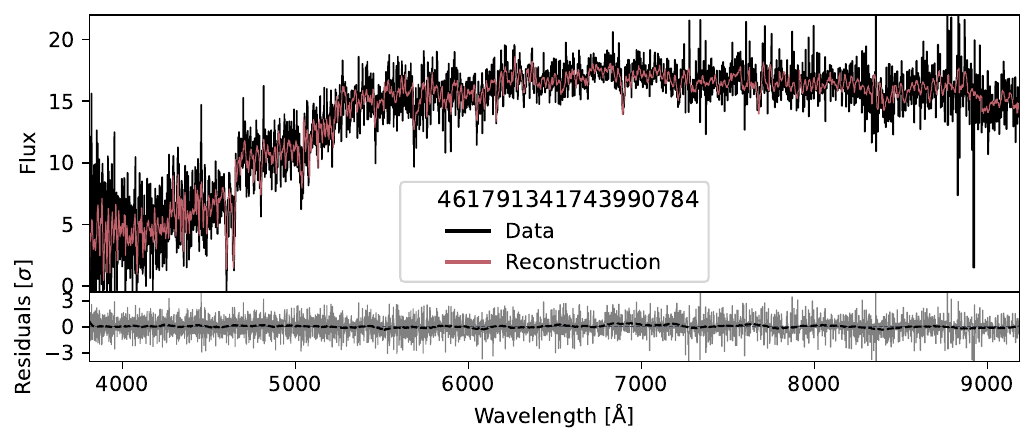}
    \caption{Reconstructed galaxy spectrum from SDSS.}
\end{subfigure}
\hfill
\begin{subfigure}{0.48\textwidth}
    \includegraphics[width=\textwidth]{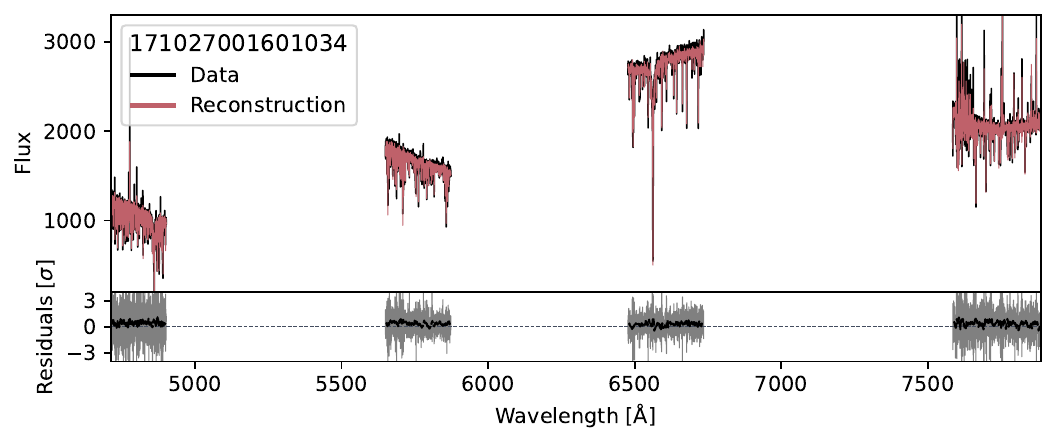}
    \caption{Reconstructed stellar spectrum from GALAH.}
\end{subfigure}
\vspace{0.5em}
\begin{subfigure}{0.48\textwidth}
    \includegraphics[width=\textwidth]{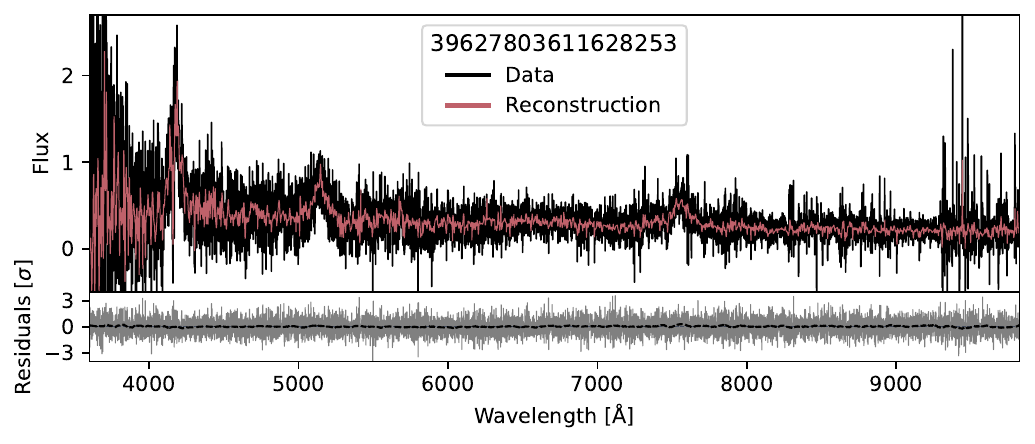}
    \caption{Reconstructed quasar spectrum from DESI.}
\end{subfigure}
\hfill
\begin{subfigure}{0.48\textwidth}
    \includegraphics[width=\textwidth]{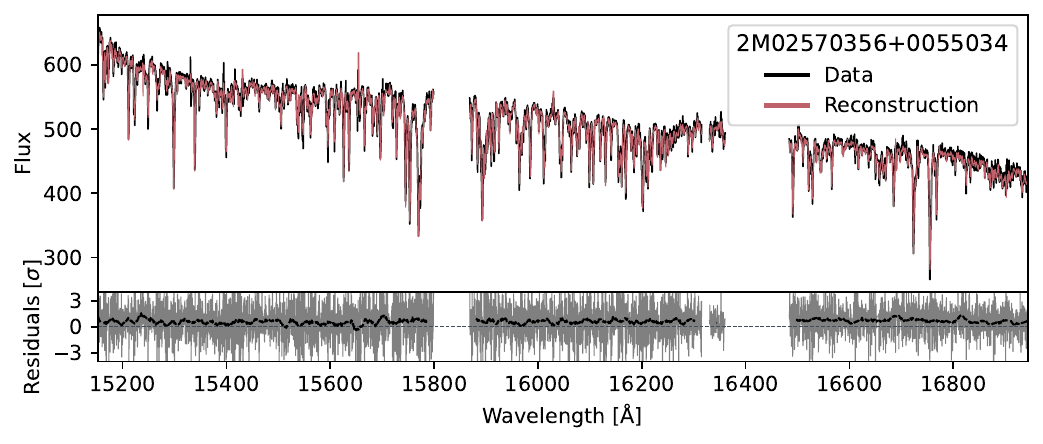}
    \caption{Reconstructed stellar spectrum from APOGEE.}
\end{subfigure}
  \caption{Examples of reconstructed spectra from the model for three different object types and four different surveys. Each subfigure represents a different dataset. The top panels show the original spectrum in black and the reconstructed spectrum in red. The bottom panels show the residuals, with raw values in grey and a smoothed version in black. Figures are best viewed zoomed in.}
\label{fig:reconstruction_example}
\vspace{-0.5cm}
\end{figure}
We show examples of autoencoded spectra for each of our four datasets in Figure \ref{fig:reconstruction_example}.
These datasets span multiple orders of magnitude in flux and exhibit a wide variety of physical phenomena and spectral features, yet a single model with our design is able to reconstruct them all.
Because the tokenizer operates directly on native wavelength grids, it naturally captures information across multiple scales---ranging from broad continuum shapes to narrow absorption and emission lines---without requiring any dataset-specific preprocessing.

\paragraph{Structure of the latent space}

\begin{figure}[t]
\centering
\includegraphics[width=0.9\textwidth]{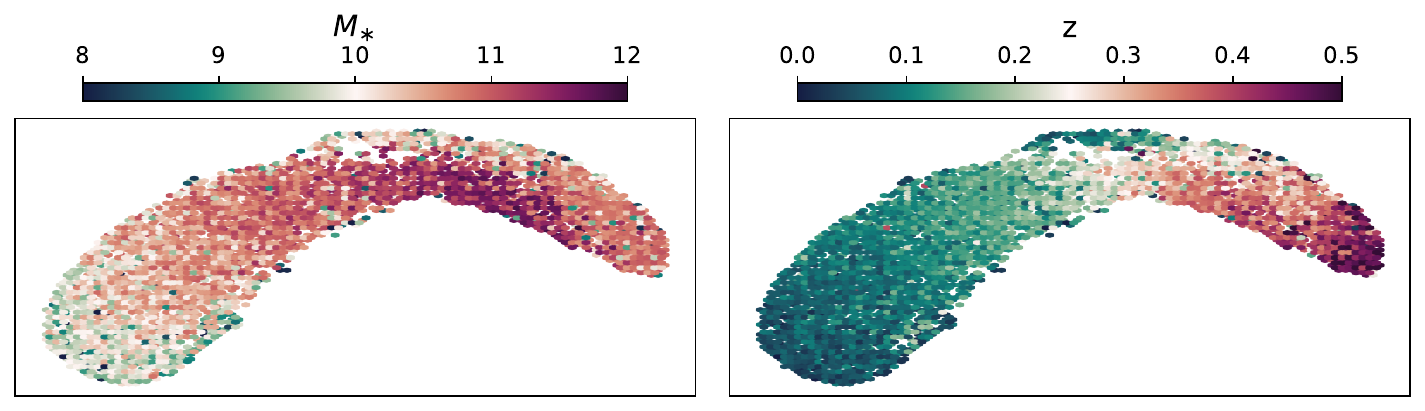}
  \caption{UMAP visualization of mean-pooled embeddings for 10,000 SDSS galaxies. The x and y axes are arbitrary. The panels are coloured by stellar mass (\textbf{left}) and redshift (\textbf{right}), and show that the embeddings have strong correlations with these physical properties.}
\label{fig:umap}
\vspace{-0.5cm}
\end{figure}

Following \cite{Melchior2022}, we use UMAP to examine the structure of the learned embedding space in Figure \ref{fig:umap}.
Here we perform this visualization for 10,000 randomly selected SDSS galaxies with no redshift warnings or plate quality issues.
We then apply UMAP to the mean-pooled embeddings%
\footnote{%
This collapses the wavelength-dependent information in the embeddings, with expected loss of intra-spectrum details;
however, this is sufficient for our goal of performing a qualitative analysis of the overall structure of the embedding space.
}
to reduce the dimensionality to 2D for visualization.
When we color the points by either their stellar mass, which is obtained from the FIREFLY spectral fitting code with the ELODIE library \citep{prugniel2001elodie, comparat2019firefly}, or by redshift, we see clear color gradients in the embedding space, indicating that the embeddings are highly correlated with these physical properties.
We again emphasize that the upstream ViT model that produced these embeddings was trained without any notion of downstream ``derived'' labels; the only data it has seen are raw spectra, and the visible structure is emergent.


\paragraph{Object classification}
We evaluate our model on the task of object classification, where the goal is to classify spectra into different object types (galaxies, stars, quasars).
We train a lightweight adaptation module on top of the embeddings from our (frozen) pretrained universal spectrum tokenizer.
In Table \ref{tab:desi_classification} we show that this achieves competitive performance in comparison to a strong task-specific baseline from \cite{zhong2024gasnet2}.
We note that our adaptation module is generic with respect to the data and task; in fact, we use the exact same architecture for the physical parameter estimation task below.
Furthermore, because the embeddings from our universal tokenizer are already informative, this adaptation module is extremely lightweight is capable of achieving strong performance with only a fraction of the training time as compared to a from-scratch model.
\paragraph{Physical parameter estimation}
\begin{wraptable}[20]{r}{0.53\textwidth}
    \vspace{-0.45cm}
    \centering
    \footnotesize
    \caption{\footnotesize Performance on object classification from DESI spectra. The results are reported as accuracy (percentage of correctly classified objects). \textbf{Higher is better.}}
    \label{tab:desi_classification}
    \begin{tabular}{lcccc}
    \toprule
      \textbf{Model} & \textbf{Galaxy} & \textbf{QSO} & \textbf{Star} & \textbf{Average} \\
    \midrule
      \cite{zhong2024gasnet2} & 93\% & 99\% & 98\% & 96\% \\
      Ours w/ adaptation & 94\% & 97\% & 98\% & 96\% \\
    \bottomrule
    \end{tabular}
    \vspace{0.2cm}
    \centering
    \footnotesize
      \caption{\footnotesize Performance on stellar parameter estimation from APOGEE spectra. $^\ast$ indicates errors expressed as $\sigma$ of residuals, and $^\dagger$ indicates $\sigma^{\rm MAD}$ of residuals, in the respective units. \textbf{Lower is better.}}
    \label{tab:apogee_regression}
    \begin{tabular}{lccc}
    \toprule
      \textbf{Model} & \textbf{log~g} & $\mathbf{T_{\rm eff}}$ & \textbf{[Fe/H]} \\
      \midrule
      \cite{casey2016cannon2}$^\ast$ & 0.07~dex & 38~K & 0.03~dex \\
      \cite{Leung2018}$^\dagger$ & 0.05~dex & 30~K & 0.02~dex \\
      \cite{olney2020apogeenet}$^\ast$ & 0.15~dex & 100~K & 0.07~dex \\
      Ours w/ adaptation$^\dagger$ & 0.07~dex & 23~K & 0.02~dex \\
    \bottomrule
    \end{tabular}
\end{wraptable}
We train an adaptation module on top of our pretrained universal spectrum tokenizer to regress physical parameters---effective temperature ($T_{\rm eff}$), surface gravity ($\log g$), and metallicity ([Fe/H])---from APOGEE stellar spectra.
We use ASPCAP estimates \citep{GarciaPerez2016} as the ground truth labels for training and evaluation; typical uncertainties from the pipeline are $\sigma_{T_{\rm eff}} = 2\%$ (${\sim}100~{\rm K}$ for a $5000~{\rm K}$ star), $\sigma_{\log g} = 0.1$~dex, and $\sigma_{\rm [Fe/H]} = 0.05$~dex\footnote{dex is used to denote an order of magnitude. Two numbers that differ by a factor of 10 differ by 1 dex, those that differ by a factor of 100 differ by 2 dex, and so on.}.
Following \cite{Leung2018}, we report performance as scaled median absolute deviation $\sigma^{\rm MAD} \approx 1.4826~{\rm MAD}$, a robust measure of dispersion. 
We show in Table \ref{tab:apogee_regression} that the performance of an adaptation module on top of our tokenizer is competitive against several well-known strong baselines, and well within uncertainties of the ASPCAP pipeline.






\section{Conclusion}

We have introduced the first universal spectral tokenizer capable of learning aligned representations across heterogeneous astronomical surveys without resampling or homogenization.
The model produces physically meaningful embeddings that can be efficiently adapted to diverse downstream tasks, achieving competitive performance with task-specific baselines.
Beyond enabling cross-survey analyses and knowledge transfer between previously isolated spectral domains, our approach provides a general framework for unifying highly heterogeneous sequential data, such time series, in a domain-agnostic and scalable fashion.

\begin{ack}

The authors would like to acknowledge the Center for Computational Astrophysics at the
Flatiron Institute for hospitality while a portion of this work was carried out. 
In addition, the data used in this work are hosted on equipment supported by the Scientific Computing Core at the Flatiron Institute, a division of the Simons Foundation. 
JS is supported by the Natural Sciences and Engineering Research Council of Canada (NSERC), funding reference number 587652, and by the Citadel GQS PhD Fellowship in Physics.
Polymathic AI and SH gratefully acknowledge the support provided by Schmidt Sciences and Simons Foundation. 

\vspace{0.4cm}

Funding for the Sloan Digital Sky 
Survey IV has been provided by the 
Alfred P. Sloan Foundation, the U.S. 
Department of Energy Office of 
Science, and the Participating 
Institutions. 

SDSS-IV acknowledges support and 
resources from the Center for High 
Performance Computing  at the 
University of Utah. The SDSS 
website is www.sdss4.org.

SDSS-IV is managed by the 
Astrophysical Research Consortium 
for the Participating Institutions 
of the SDSS Collaboration including 
the Brazilian Participation Group, 
the Carnegie Institution for Science, 
Carnegie Mellon University, Center for 
Astrophysics | Harvard \& 
Smithsonian, the Chilean Participation 
Group, the French Participation Group, 
Instituto de Astrof\'isica de 
Canarias, The Johns Hopkins 
University, Kavli Institute for the 
Physics and Mathematics of the 
Universe (IPMU) / University of 
Tokyo, the Korean Participation Group, 
Lawrence Berkeley National Laboratory, 
Leibniz Institut f\"ur Astrophysik 
Potsdam (AIP),  Max-Planck-Institut 
f\"ur Astronomie (MPIA Heidelberg), 
Max-Planck-Institut f\"ur 
Astrophysik (MPA Garching), 
Max-Planck-Institut f\"ur 
Extraterrestrische Physik (MPE), 
National Astronomical Observatories of 
China, New Mexico State University, 
New York University, University of 
Notre Dame, Observat\'ario 
Nacional / MCTI, The Ohio State 
University, Pennsylvania State 
University, Shanghai 
Astronomical Observatory, United 
Kingdom Participation Group, 
Universidad Nacional Aut\'onoma 
de M\'exico, University of Arizona, 
University of Colorado Boulder, 
University of Oxford, University of 
Portsmouth, University of Utah, 
University of Virginia, University 
of Washington, University of 
Wisconsin, Vanderbilt University, 
and Yale University.

This research used data obtained with the Dark Energy Spectroscopic Instrument (DESI). DESI construction and operations is managed by the Lawrence Berkeley National Laboratory. This material is based upon work supported by the U.S. Department of Energy, Office of Science, Office of High-Energy Physics, under Contract No. DE–AC02–05CH11231, and by the National Energy Research Scientific Computing Center, a DOE Office of Science User Facility under the same contract. Additional support for DESI was provided by the U.S. National Science Foundation (NSF), Division of Astronomical Sciences under Contract No. AST-0950945 to the NSF’s National Optical-Infrared Astronomy Research Laboratory; the Science and Technology Facilities Council of the United Kingdom; the Gordon and Betty Moore Foundation; the Heising-Simons Foundation; the French Alternative Energies and Atomic Energy Commission (CEA); the National Council of Humanities, Science and Technology of Mexico (CONAHCYT); the Ministry of Science and Innovation of Spain (MICINN), and by the DESI Member Institutions: www.desi.lbl.gov/collaborating-institutions. The DESI collaboration is honored to be permitted to conduct scientific research on I’oligam Du’ag (Kitt Peak), a mountain with particular significance to the Tohono O’odham Nation. Any opinions, findings, and conclusions or recommendations expressed in this material are those of the author(s) and do not necessarily reflect the views of the U.S. National Science Foundation, the U.S. Department of Energy, or any of the listed funding agencies.

This work made use of the Third Data Release of the GALAH Survey (Buder et al. 2021). The GALAH Survey is based on data acquired through the Australian Astronomical Observatory, under programs: A/2013B/13 (The GALAH pilot survey); A/2014A/25, A/2015A/19, A2017A/18 (The GALAH survey phase 1); A2018A/18 (Open clusters with HERMES); A2019A/1 (Hierarchical star formation in Ori OB1); A2019A/15 (The GALAH survey phase 2); A/2015B/19, A/2016A/22, A/2016B/10, A/2017B/16, A/2018B/15 (The HERMES-TESS program); and A/2015A/3, A/2015B/1, A/2015B/19, A/2016A/22, A/2016B/12, A/2017A/14 (The HERMES K2-follow-up program). We acknowledge the traditional owners of the land on which the AAT stands, the Gamilaraay people, and pay our respects to elders past and present. This paper includes data that has been provided by AAO Data Central (datacentral.org.au).

\end{ack}

\bibliographystyle{plainnat}
\bibliography{references,references_extra}


\appendix

\end{CJK*}
\end{document}